\newif\ifAnon\Anonfalse
\newif\ifDraft\Draftfalse
\newif\ifFinal\Finaltrue

\ifFinal\Anonfalse\Draftfalse\fi

\documentclass[conference]{IEEEtran}

\def\BibTeX{{\rm B\kern-.05em{\sc i\kern-.025em b}\kern-.08em
    T\kern-.1667em\lower.7ex\hbox{E}\kern-.125emX}}

\usepackage{CAMLsec}
\usepackage{colortbl}
\usepackage{enumitem}
\usepackage{subcaption}
\usepackage{framed}
\usepackage{balance}
\usepackage{makecell}
\usepackage{tikz}   
\newcommand*\numcircledtikz[1]{\tikz[baseline=(char.base)]{
            \node[shape=circle,draw,inner sep=1.2pt] (char) {#1};}} 
\usepackage{mathrsfs}

\usepackage{cite}
\usepackage{amsmath,amssymb,amsfonts}
\usepackage{algorithmic}
\usepackage{graphicx}
\usepackage{textcomp}
\usepackage{xcolor}
\usepackage{tabularx}

\usepackage{hyperref}
\usepackage{fancyhdr}
\usepackage[hyphens]{url}

\newtheorem{theorem}{Theorem}
\newtheorem{definition}{Definition}

\def\BibTeX{{\rm B\kern-.05em{\sc i\kern-.025em b}\kern-.08em
    T\kern-.1667em\lower.7ex\hbox{E}\kern-.125emX}}

\title{\shmds: Adversarial Resilient Hardware Malware Detectors through Voltage Over-scaling}

\author{\IEEEauthorblockN{Md Shohidul Islam\IEEEauthorrefmark{1}\IEEEauthorrefmark{4}, Ihsen Alouani\IEEEauthorrefmark{3}, Khaled N. Khasawneh\IEEEauthorrefmark{1}}
\IEEEauthorblockA{
\textit{\IEEEauthorrefmark{1}ECE Dept., George Mason University, Fairfax, VA, USA} \\
\textit{\IEEEauthorrefmark{4}CSE Dept., Dhaka University of Engineering \& Technology, Gazipur, Bangladesh} \\
\textit{\IEEEauthorrefmark{3}IEMN CNRS-UMR8520, Universite Polytechnique Hauts-De-France,  Valenciennes, France} \\
Email: \{mislam20, kkhasawn\}@gmu.edu, ihsen.alouani@uphf.fr
}
}

\begin{document}
\maketitle
\pagestyle{plain}

\begin{abstract}

 Machine learning-based hardware malware detectors (HMDs) offer a potential game changing advantage in defending systems against malware. 
 However, HMDs suffer from adversarial attacks, can be effectively reverse-engineered and subsequently be evaded, allowing malware to hide from detection. We address this issue by proposing a novel HMDs (\shmds) through approximate computing, which makes HMDs' inference computation-stochastic, thereby making HMDs resilient against adversarial evasion attacks. Specifically, we propose to leverage voltage overscaling to induce stochastic computation in the HMDs model. We show that such a technique makes HMDs more resilient to both black-box adversarial attack scenarios, i.e., reverse-engineering and transferability. Our experimental results demonstrate that \shmds offer effective defense against adversarial attacks along with by-product power savings, without requiring any changes to the hardware/software nor to the HMDs' model, i.e., no retraining or fine tuning is needed. Moreover, based on recent results in probably approximately correct (PAC) learnability theory, we show that \shmds are provably more difficult to reverse engineer.

\end{abstract}


\maketitle

\section{Introduction}
Computing systems are under continuous attacks by increasingly motivated and sophisticated adversaries. These attackers exploit vulnerabilities to compromise systems and deploy malware. Although significant effort continues to be directed at making systems more resilient to attacks, the number of exploitable vulnerabilities is overwhelming. While preventing compromise is difficult, signature based static analysis techniques can be easily bypassed using metamorphic/polymorphic malware or zero-day exploits since their signatures have not yet been encountered~\cite{moser-07}. On the other hand, dynamic detection techniques can detect unseen signatures since they monitor the behavior of the program. However, the complexity and difficulty of continuous dynamic monitoring have traditionally limited its use due to constrained resources.

Against this backdrop, several research studies proposed using Hardware Malware Detectors (HMDs) to make the continuous dynamic monitoring resource-efficient~\cite{demme2013feasibility,khasawneh-15,ozsoyhardware,KazdagliHRT16,patel2017analyzing,khasawneh2018ensemblehmd}. Specifically, these studies showed that malware can be classified as a computational anomaly using low-level hardware features, such as instructions traces.
Moreover, it appears that the industry started to show interest in using HMDs too; SnapDragon processor from Qualcomm appears to be using hardware features to detect malware, but the technical details are not published~\cite{qualcomm-16}. Therefore, HMDs can offer a significant advantage to defend against malware attacks because they can be `always on' with small-to-no impact on performance~\cite{ozsoyhardware,patel2017analyzing}.

As HMDs showed potential defense effectiveness, it is natural to expect that attackers attempt to find adaptive ways to evade detection. As a consequence, it was shown that attackers can adapt malware to continue to operate while avoiding detection by HMDs~\cite{khasawneh2017rhmd,dinakarrao2019adversarial}. These attacks assume black-box access to the HMDs; the attacker can only query the victim HMD and observe the output. Therefore, the attack consists of two steps: (1) Reverse-engineering the victim HMD to create a proxy model and (2) Developing evasive malware based on the proxy model to bypass detection while preserving the malware's intended functionality. 

Among the proliferation of defense proposals against adversarial attacks in the computer vision domain, randomization-based defenses are shown to be promising for improving model robustness against adversarial attacks~\cite{khasawneh2017rhmd,  sengupta2019mtdeep, huang2019model, liu2018towards, wang2019protecting}. 
The main idea is to make the detection model non-deterministic. In particular, variations of a non-deterministic model lead to random input gradients and therefore perplexes the adversary, which makes reverse-engineering the model substantially more difficult and thus complicates evasion. As a result, multiple randomization techniques have been proposed, switching between multiple diverse models~\cite{khasawneh2017rhmd,wang2019protecting,sengupta2019mtdeep}.
However, these proposed techniques require substantial changes to the architecture or retraining/fine-tuning procedures, which involve significant additional overhead, making it more challenging to build secure models/HMDs under power and computational resource constraints.

Today's electronic devices are reaching the physical limits of CMOS technology. However, the overall energy consumption of computing systems is still rapidly growing to cope with the ever-increasing performance requirements~\cite{DandT_survey}.
On the other hand, a wide range of computing domains and applications, such as recognition, mining, and synthesis (RMS), feature an intrinsic error-tolerance property~\cite{kaushik_roy}. For this reason, in such applications, a migration in the design methodology towards Approximate Computing (AC) rather than unnecessarily exact computing is witnessed. AC is a design paradigm that seeks a balance between accuracy, on one hand, and power consumption, on the other hand. 
Using AC, conventional digital circuits' power constraint can be overcome by compromising accuracy. While the performance and power consumption challenges drive the state-of-the-art primary use of AC, \emph{a novel utility} of this paradigm is proven in this work in addition to its initial advantage.


In this paper, we propose \shmds, which are HMDs that unprecedentedly utilize AC to defend against adversarial attacks. In particular, we leverage voltage overscaling (VOS) to induce stochastic computations in HMD's model during inference. 
From the security perspective, stochastic computations during inference can be utilized by HMDs to make their model non-deterministic, i.e., the model's decision boundaries will be stochastic. This aspect allows HMDs to resist reverse-engineering and to complicate evasion by changing the decision boundary at runtime. From the performance perspective, VOS offers a by-product  power savings while trading it with small accuracy loss for the sake of security. 

The key contributions of this paper are as follows:
\begin{itemize}
    \item To the best of our knowledge, this is the first work that exploits VOS for security. In particular, our work is the first to show that computational faults can be used as a defensive technique against adversarial attacks. 
    \item \shmds enhance the robustness of HMDs against black-box attacks. The stochastic computation induced by VOS makes reverse engineering harder and thereby reduces the transferability of evasive malware.
    \item We evaluate the implementation overhead of both hardware and software-based \shmds. Our results demonstrate that \shmds significantly reduce power consumption compared to an unsecured HMD and offer area, power, and performance savings compared to state-of-the-art HMD defense, namely RHMD~\cite{khasawneh2017rhmd}. Thus, \shmds offer a practical security solution, especially for energy-constrained devices. 
    \item Lastly, PAC learnability theory was utilized to formally show the advantage of \shmds. 
\end{itemize}





\section{Background}\label{sec:background}

\subsection{Adversarial attacks on machine learning}
Machine learning (ML) algorithms are shown to be vulnerable to adversarial attacks, i.e., carefully crafted noise that compromises classifiers' integrity. Main threat categories of adversarial attacks include \textit{poisoning attacks} and \textit{evasion attacks}. In poisoning attack, sophisticated attackers inject specially crafted malicious data or attack points in the \textit{training dataset}, exploiting the fact that ML models are often trained with outside world data or user-provided data; such deliberate poisoning of the \textit{training data} manipulates the trained model behavior during test time as per the attackers' goal~\cite{demontis2019adversarial, steinhardt2017certified, jagielski2018manipulating }. In contrast, in evasion attacks, adversaries perturb \textit{test data}. For malware detection, evasion attacks confuse the model and produce misclassification, thereby enabling the evasive malware to perform their intended malicious activity while behaving like benign applications \cite{chen2017adversarial}. In this work, we consider a set of powerful \textit{evasion attacks} and propose a novel defense against them.

\subsection{Approximate Computing through VOS}
\label{sec:vos}

A wide range of emerging applications feature an inherent fault-tolerance property~\cite{kaushik_roy}. This property has led to the emerging of AC, which leverages circuit, architecture and software level techniques to reduce power consumption and resource utilization with a tolerable accuracy loss. While the state-of-the-art main use of AC is performance and power-oriented, \emph{a novel utility of this paradigm is shown in this work}.


In this paper, we use a circuit level approximation, specifically VOS~\cite{volt}, to harden ML classifiers against adversarial attacks. VOS is different from conventional dynamic voltage and frequency scaling since we do not accordingly scale the clock frequency, thereby intentionally causing the critical paths to randomly violate the clock period~\cite{volt} while reducing the supply voltage. 

\textbf{Impact of VOS:} While the transistor size shrinks with the new device generations, the effect of process variation becomes more critical from a digital design perspective. These variations are mainly created by factors such as imperfection of the manufacturing process, random dopant fluctuation, and variation in the gate oxide thickness. With reduced transistors dimensions, the standard deviation of threshold voltage variation ($\Delta \mathrm{V_t}$) increases since it is proportional to the square root of the device area \cite{volta_ref8}, as shown in Equation \ref{eq:delta}




\begin{equation}
\label{eq:delta}
    \sigma_{\Delta \mathrm{V_t}}=\frac{\mathrm{A}_{\Delta \mathrm{V_t}}}{\sqrt{\mathrm{WL}}}
\end{equation}

where $W$ and $L$ are the width and the length of the device, respectively, and $\mathrm{A}_{\Delta \mathrm{V_t}}$ is characterizing matching parameter for any given process. This variation in $V_t$ will have a direct impact on the delay of a CMOS gate which can be approximated using the following equation \cite{volta_ref8}:

\begin{equation}
\label{eq:gate}
    d_{g a t e} \propto \frac{V_{D D}}{\beta\left(V_{D D}-V_{t}\right)^{\alpha}}
\end{equation}

where $\alpha$ and $\beta$ are fitting parameters for a given gate in a given process.  As shown in Equation \ref{eq:gate}, scaling down the supply voltage ($V_{dd}$) from the nominal voltage results in slowing down signals propagation and thereby creating a timing overhead. If this process is \emph{not} accompanied with a corresponding frequency scaling, errors may occur within the circuit results: this is the case of VOS. These timing violation-induced errors are \textbf{stochastic} by nature due to the operating and manufacturing variability. However, their magnitude is statistically shaped by the voltage level: the more aggressive the VOS, the higher the error rate. 

\section{Dataset \& feature collection}\label{sec:data}
The dataset consists of 3000 malware and 600 benign programs. The malware programs were downloaded from the Zoo malware database (a.k.a MalwareDB)~\cite{new-malware} and include five types of malware, which are backdoors, rogues, password stealers, trojans, and worms. The benign programs consisted of a wide variety of applications, including browsers, text editing tools, system programs, and CPU performance benchmarks. The dataset was divided evenly into 4-folds, which are \textit{victim training\_1}, \textit{victim training\_2}, \textit{attacker training}, and \textit{testing}. In particular, the \textit{victim training\_1} and \textit{victim training\_2} combined are called \textit{victim training} and will be used to train our baseline HMDs, \textit{attacker training} is used by the attacker to perform reverse-engineering of the victim HMD and build a substitute/proxy model, and \textit{testing} to create evasive malware and in performance analysis of both HMDs and \shmds. In addition, we use 4-fold cross-validation in our experiments to get more accurate results. Furthermore, we made sure that the malware types as well as the benign application types were distributed evenly and randomly across the folds to ensure that the datasets are not biased.

%

To collect low-level hardware features dynamically, we used Intel's Pin instrumentation tool~\cite{Pin} running on an isolated machine that is running Windows 7. All security and firewall services were disabled to ensure that malware runs freely. Furthermore, we collect a trace for each executable until it executes 5,000 system calls or 15 million committed instructions. The extracted features from the traces are based on the frequency of executed instruction categories. \textcolor{black}{The instruction categories are based on Intel's sub-grouping of instructions~\cite{intel_sdm}}. The trace collection and features extraction are modeled after the RHMD study~\cite{khasawneh2017rhmd} to make sure that the data is sufficiently large to establish the feasibility and provide trustworthy experimental results.

Recently, it has been shown that hardware features collected through hardware performance counters (HPCs) are not reliable to be used in security application due to their non-determinism~\cite{das2019sok}. In this work, we do \textbf{not} use HPCs, and we make sure that our feature collection framework is deterministic; we get the exact same trace as the expected trace every run when we supply the same input. We manually verified this by running samples of programs multiple times, using different machines and using a virtual machine.

\section{Adversarial attacks on HMDs: The threat model}
\begin{figure}[]
\includegraphics[width=0.9\linewidth]{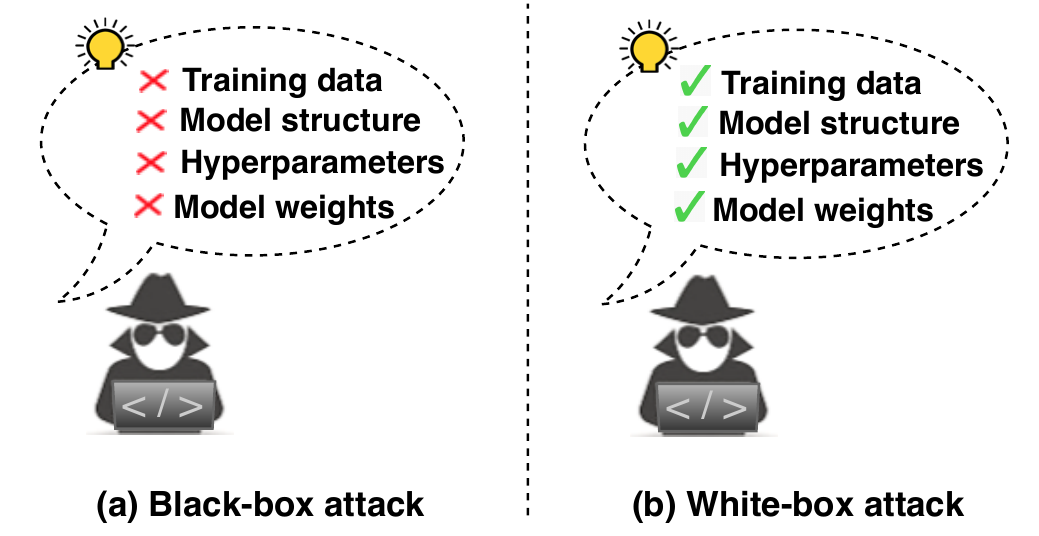} 
\centering
\caption{Attacker knowledge for black-box attack compared to a white-box attack scenario.}
\label{fig:knowledge}
\end{figure}

We assume a black-box adversarial attack scenario based on the adversary's knowledge of the victim HMD internal model, as shown in Figure~\ref{fig:knowledge}. Therefore, in this section, we will describe our baseline (HMD). Then we will
follow by explaining the black-box adversarial attack and show that the
current HMDs are vulnerable to this threat model.

\subsection{Baseline HMD}\label{subsec:victimHMD}
HMDs use low-level features to classify malware as a computational anomaly at run-time. 
Therefore, we trained our baseline HMD on a multi-layer perception (MLP) neural network using the features described in Section~\ref{sec:data}. The MLP consists of a single hidden layer that has the number of neurons equal to the number of input features (50 neurons). We used the Rectified Linear Unit (ReLU) as an activation function. The rationale for selecting neural networks to train our baseline HMD is that it has the highest performance in detecting malware~\cite{ozsoyhardware}. 

The \textit{baseline training} set was used to train our baseline HMD and the \textit{testing} data set was used to evaluate its detection performance. Table~\ref{tbl:vhmd_per} shows the baseline HMD detection performance 
using the following metrics: \textit{accuracy} (how many programs the baseline HMD correctly labeled out of all programs), \textit{sensitivity} (how many malware the baseline HMD correctly classified out of all programs that were labeled as malware), \textit{specificity} (how many benign programs the baseline HMD correctly classified out of all programs that were labeled as benign), \textit{precision} (how many of those who were labeled as malware are actually malware), and \textit{F1-score} (the harmonic mean of the precision and sensitivity). The results show that the baseline HMD achieves high performance in detecting malware across all performance metrics.

\begin{table}[h!]
\centering
\small
\caption{Baseline HMD detection performance}
\label{tbl:vhmd_per}
\begin{tabular}{ccccc}
 \hline
                                        
             \begin{tabular}[c]{@{}c@{}}\textbf{Accuracy}\end{tabular}  &
             \begin{tabular}[c]{@{}c@{}}\textbf{Sensitivity}\end{tabular}  &
             \begin{tabular}[c]{@{}c@{}}\textbf{Specificity}\end{tabular}  &
              \begin{tabular}[c]{@{}c@{}}\textbf{Precision}\end{tabular}  &
             \begin{tabular}[c]{@{}c@{}}\textbf{F1-Score}\end{tabular}   \\ \hline
  92.7\% &  92.6\%    & 93.0\%     &  98.0\% &  95.2\%    \\ \hline
\end{tabular}
\smallskip
\end{table} 

Note that this detector will be used throughout this study; (1) to demonstrate that current HMDs are vulnerable to adversarial attacks (Section~\ref{sec:BBA}), and (2) to construct \shmds that are resilient to adversarial attacks (Section~\ref{sec:approxHMD}).


\subsection{Black-box attack: }\label{sec:BBA}
\begin{figure}[]
\includegraphics[width=\linewidth]{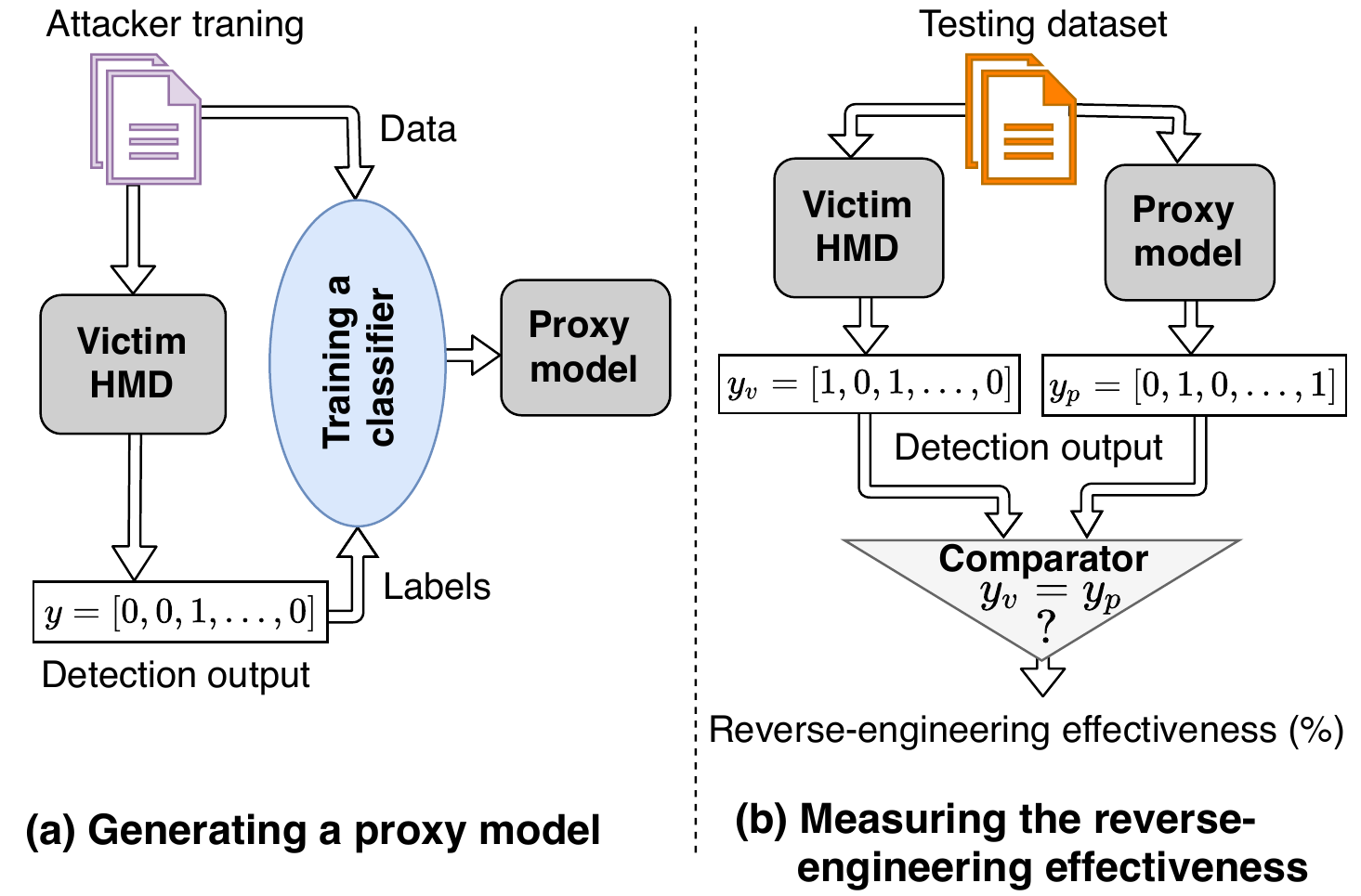} 
\centering
\caption{Reverse-engineering methodology} 
\label{fig:overview-reverse-engineering}
\end{figure}

 The adversarial attack model starts with the adversary attempting to reverse engineer the victim HMD, 
 then utilizing the reverse-engineered HMD, i.e., proxy model, to generate evasive malware examples that can bypass the detection. 

\noindent
\textbf{\textit{Reverse Engineering}.} In black-box attacks, the attacker has access only to the input/output of the victim HMD and has no information about its internal architecture, e.g., structure, weights, hyperparameters, or training data. 
However, the HMD internal model is necessary for the attacker to be able to develop evasive malware methodically; otherwise, the problem becomes NP-Hard~\cite{Voro-Li14}. Therefore, in such a setting, the attacker utilizes the observed inputs/outputs of the victim HMD to reverse-engineer it and train a proxy model~\cite{khasawneh2017rhmd}. An overview of the reverse-engineering methodology is shown in Figure~\ref{fig:overview-reverse-engineering}. Specifically, the attacker prepares a data set that contains both malware and benign programs and uses the victim HMD to label these inputs. 
Thus, a new data set that is labeled by the victim HMD output will be constructed. Finally, the newly constructed data set will be used to train a proxy model. Subsequently, to evaluate reverse-engineering effectiveness, we compare the proxy model performance against the victim HMD. In particular, the attacker has to prepare a testing set (separate from the set used to reverse-engineer the victim HMD) and query both the substitute model and the victim HMD using the testing set. The percentage of similar classifications made by the two detectors represents reverse-engineering effectiveness.

We evaluated the effectiveness of the black-box attack (reveres-engineering) on our baseline HMD. For this experiment, we used the \textit{attacker training} set to train a proxy model and the \textit{testing} set to evaluate our reveres-engineering effectiveness. In addition, we used an MLP neural network with the same structure (number of hidden layers and number of neurons) as the baseline HMD to train the proxy model. Our result shows that we were able to effectively reverse-engineer the \emph{baseline HMD}, with less than \textbf{1\%} error. This result demonstrates that current HMDs can be reverse-engineered.

\noindent
\textbf{\textit{Evasive samples generation}.}
After reverse-engineering the victim HMD, the attacker's goal is to utilize the knowledge of the proxy model, i.e., reverse-engineered HMD, to systematically create evasive malware. In particular, the attacker has to start by identifying the features that can be used to create evasive malware and then embed the identified features in the malware without changing their intended functionality.

Firstly, to identify the features that can be used to create evasive malware, we employ a slightly modified version of the Fast-Gradient Sign Method (FGSM), which is widely employed in image processing~\cite{goodfellow2014explaining}, to identify the features that we can use to create an evasive malware. Therefore, we assumed that $\theta$ is the parameter of the victim HMD, $x$ is the input, $y$ is the output for a given input, and $L(\theta, x, y)$ is the cost function. Then the changes to the necessary features are determined based on the cost function gradient. In contrast to FGSM method, due to the nature of the HMD's features (follow a positive continuous distribution, i.e., cannot have a negative value) and the nature of the evasion strategy (based on adding instruction), we only allow positive changes to the features ($x$). Basically, using this method, we were able to identify the features that, by adding a small random perturbation to their current value, would result in raising the HMD confidence that the input is a benign program. Note that the features in our case are instructions (Section~\ref{sec:data}).

After identifying the features, e.g., instructions, we need to systematically embed them in the malware to transform them into evasive malware. Our malware set consists of actual malware binaries; we did not have access to their source code and were not able to decompile them since they were obfuscated. Therefore, we follow a similar approach to the one proposed in~\cite{khasawneh2017rhmd}. In particular, for each malware program, we constructed a Dynamic Control Flow Graph (DCFG) using Intel's Pin tool~\cite{Pin}. Then we added instructions, identified from the previous step, to each basic block of the DCFG. Furthermore, when we add each instruction, we ensure that they do not affect the state of the program to preserve the malware's intended functionality. 
For example, if we are adding an \verb|add| instruction, we add \textit{zero} to any register.

\begin{figure}[]
\includegraphics[width=\linewidth]{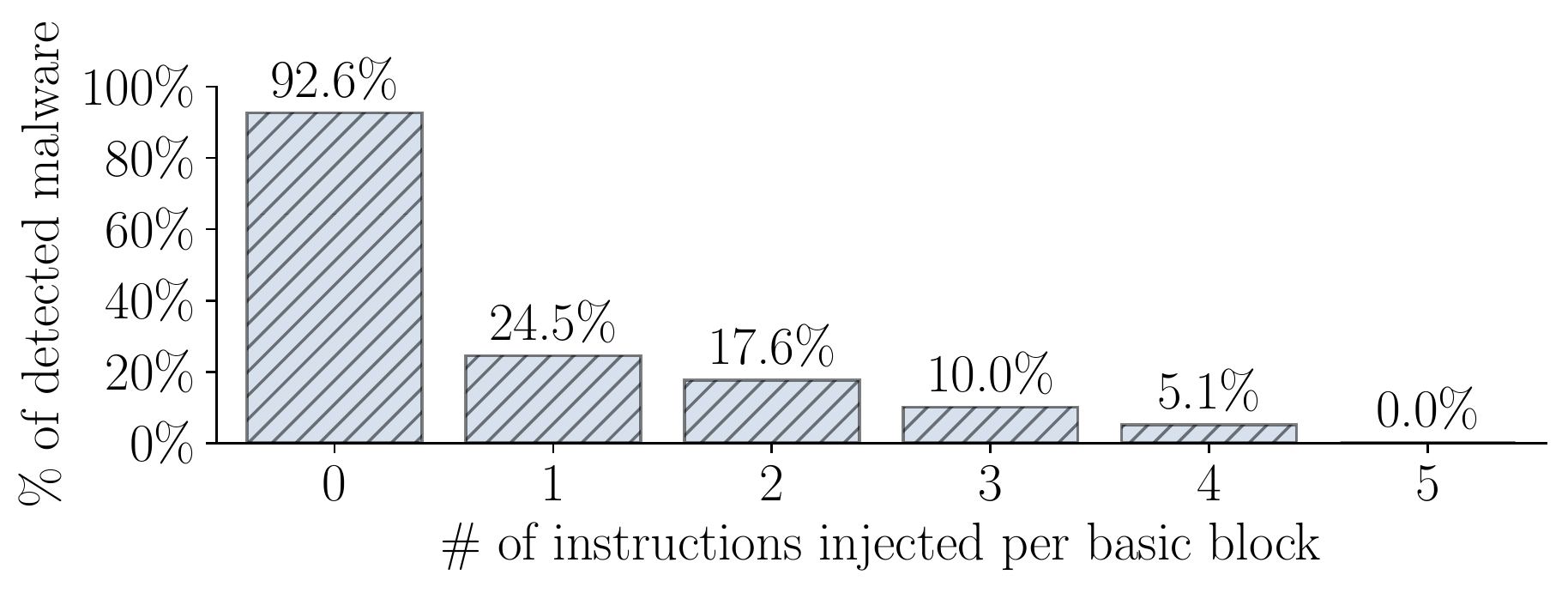} 
\centering
\caption{Transferability: black-box attack effectiveness on current HMDs} 
\label{fig:wb_baseline}
\end{figure}

\noindent
\textbf{\textit{Transferability}.} 
 To evaluate the performance of black-box attacks on our baseline HMD, we measure the transferability of evasive malware created using the reverse-engineered HMD. Specifically, transferability represents the percentage of evasive malware created using the reverse-engineered HMD, i.e., proxy model, that can also evade the baseline HMD, i.e., the victim HMD. In this experiment, we used the \textit{testing} set to create evasive malware samples. Figure~\ref{fig:wb_baseline} shows that by injecting one instruction per basic block, the malware detection accuracy of the victim HMD, i.e., the baseline HMD, drops from around 92\% to around 24\%. Moreover, the figure shows that the more instructions we inject per basic block, the more malware can evade detection. This result shows that current HMDs are vulnerable to black-box attacks.

\section{\shmds}\label{sec:approxHMD}
In this section, we propose a new class of adversarial evasion resilient HMDs (\shmds). \shmds perturb the inference computation of an HMD, making the HMD's decision boundaries stochastic over time. Thus, they prevent the adversary from having reliable access to the HMD's output (reverse-engineering attacks) and reduce the transferability of evasive malware built using the victim HMD's exact model. Specifically, \shmds exploit AC, which is a computing paradigm that can trade energy consumption and computing time with the accuracy of results. Traditionally, this is a significant advantage for error-resilient applications (such as deep/machine learning, big data analytics, and signal processing) with respect to performance, power efficiency, flexibility, and cost. However, in this work, we reveal an additional advantage of using AC for a machine learning application, which is security. 

Several noise-based methods such as \cite{snP2019_certif,smooth} have been used in the computer vision community to defend against adversarial attacks.
However, these techniques add substantial performance and power overhead to be adapted; they need to query a physical noise source to guarantee the random distribution of noise and require changes to the classification model, e.g, adds noise injection layers to the model structure.  
In this paper, we use a circuit level approximation, specifically VOS~\cite{volt} as a practical source of randomness to harden HMDs against adversarial attacks. Our goal is to intentionally cause random timing violations driven by the supply voltage level~\cite{volt}. The rationales behind choosing VOS are the following:

\noindent
   \textbf{(i) Injects time-variant stochastic behavior:}
 one of the interesting properties of VOS is that the timing violations are stochastic. 
 This offers a significant advantage to defend against adversarial attacks because it offers time-variant decision boundaries, i.e., a moving target defense~\cite{khasawneh2017rhmd}. In contrast, other circuit level approximation techniques depend on reducing resource utilization and redesigning the circuit blocks accordingly~\cite{apxmul}. Therefore, their behavior is deterministic, i.e., the model's decision boundaries do not change over time.
 
\textcolor{black}{Although, from a circuit perspective, under a given voltage, and for the same random dopant fluctuation and thermal conditions, the timing violations should be systematic. Nevertheless, the stochastic error behavior is due to the following:}

\begin{enumerate}
    \item \textcolor{black}{Input dependence of the critical path: different sets of operands may lead to different critical-path lengths for a given operation. For example, in simple addition, a carry rippling to different gates depend on the operands. Thus, operands are not equal with respect to timing errors even for the same operation, within the same chip, and under the same voltage.} 
    
    \item \textcolor{black}{Masked timing violations: a timing violation for a given path does not lead systematically to error propagation. The error manifestation due to timing violation depends on both current and previous latched outputs of the combinational circuit. Thus, from a behavioral perspective, this makes the errors appear stochastically in the output.}
    
    \item \textcolor{black}{Variable thermal impact: the temperature’s impact on voltage threshold and delay varies with the voltage~\cite{Handling2006}. Hence, by considering on-chip thermal variability, timing violations are stochastically impacted by temperature.}
\end{enumerate}

    \noindent
    \textbf{(ii) Easy to deploy:} VOS is a practical source of randomness that can be easily deployed; it does not require any changes to the computing stack, e.g., hardware or software, since it only depends on scaling the supply voltage.

  \noindent
    \textbf{(iii) Reduces energy consumption:} Unlike related randomness-based defenses that imply high performance and power overhead, VOS comes with a by-product reducing energy consumption savings due to the super-linear dependence of both dynamic and leakage power on supply voltage (Section~\ref{subsec:pow_sav}). Specifically, for related randomness-based defenses, generating randomness requires a source of entropy, e.g., a random number generator (RNG)~\cite{drng}. As a result, they require $n$ RNG queries for each of the $n$ MAC operations in a convolution layer, which comes with increasing costs for deeper models. Please note that the RNG is an off-core components, i.e., shared between all CPU cores, thus, require more time and power to query compared to querying on-core resources.   
    

Even though perturbation-based defenses are not new~\cite{khasawneh2017rhmd, jia2018attriguard, dhillon2018stochastic, smooth}, our proposed approach advances the state of the art in general because it does not require any changes to the software or the hardware. Specifically: \textbf{(1)} it does not require additional training or fine-tuning of the protected model~\cite{jia2018attriguard,khasawneh2017rhmd}, \textbf{(2)} it does not require any changes to the pre-trained model~\cite{dhillon2018stochastic}, \textbf{(3)} it does not require additional computations on the input~\cite{buckman2018thermometer}, and \textbf{(4)} it provides performance and energy gains~\cite{volt}
(details are in Section~\ref{sec:related}). To the best of our knowledge, this is the first work that exploits VOS for security.

\section{Security Evaluation}\label{sec:eval}



In this section, we explore whether \shmds can increase the robustness of HMDs against adversarial attacks. Our results show that the computational faults induced by VOS help making HMDs resistant to adversarial attacks. In particular, VOS computational faults introduce noise to the black-box (reverse-engineering) attack that is bounded by a function of the VOS-induced computational faults rate. 
 
\subsection{Experimental setup}\label{subsec:shmd_exp}
 For our experimental setup, we apply VOS to the baseline HMD (Section~\ref{subsec:victimHMD}) to construct a \shmd. In particular, to model the computational stochasticity caused by the VOS, we built a stochastic fault injection tool\footnote{link omitted for blind review.} that emulates timing violations at the output of arithmetic operations, based on the error distribution model detailed earlier in Section \ref{sec:vos}. Practically, the tool injects timing violation errors that follow the distribution that matches the VOS level. Note that we did not implement any measures to detect or correct potential timing violation-induced computational faults since they are the foundation of our proposed defense. Finally, we integrated our tool to the Fast Artificial Neural Network Library (FANN)~\cite{fann} to simulate the behavior of our neural network model under VOS (the \shmd). 
 
 This detector will be used throughout the rest of this study to experimentally evaluate \shmds security (Sections~\ref{sec:shmd_bba}) and performance (Section~\ref{sec:perf}).


\subsection{Resilience against black-box attacks}\label{sec:shmd_bba}\label{sec:shmd_transf}
\begin{figure}[]
\includegraphics[width=\linewidth]{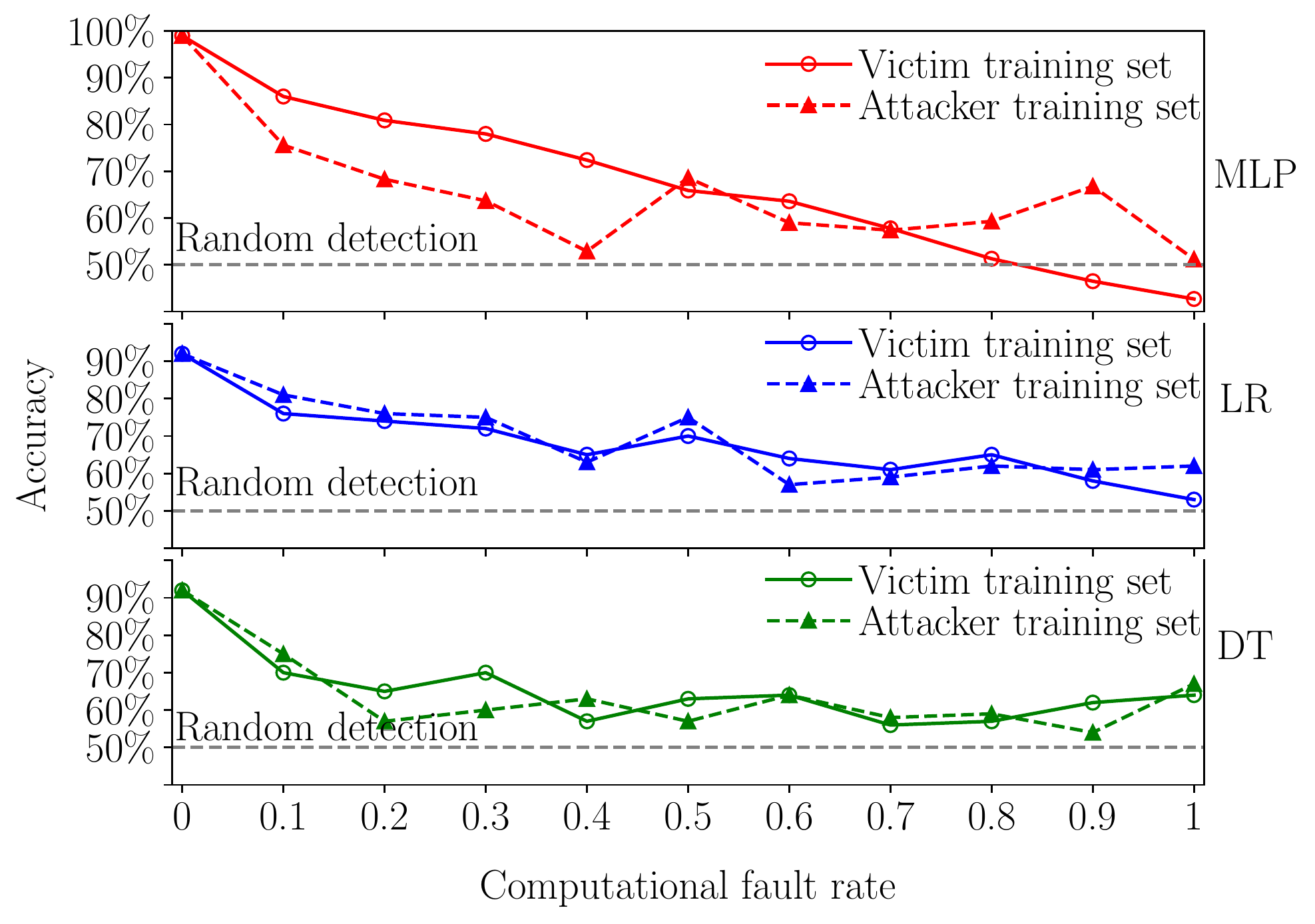} 
\centering
\caption{\shmds resilience against black-box attacks (reverse-engineering)
}
\label{fig:reverse}
\end{figure}


We first examine \shmds resilience against black-box attacks. We assume two black-box attack scenarios: (1) the attacker knows the \shmd (victim) training data; we use the \textit{victim training} set to perform the reverse-engineering, and (2) the attacker does not know the \shmd (victim) training data; we use the \textit{attacker training} set to perform the reverse-engineering. Note that the first scenario assumes a stronger attacker than the second scenario since the attacker will use the same data distribution that the victim is trained on. In addition, in both scenarios, we use the \textit{testing} set to evaluate the proxy model performance (measure the reverse-engineering effectiveness). Second, we explore the transferability of evasion binaries designed using the reverse-engineered model to the defender model. Moreover, we assume that the attacker knows the \shmd (victim) structure. Therefore, we use the same MLP neural network structure to perform the attack. In addition, we have performed the black-box attack using other machine learning algorithms: Logistic Regression (LR) and Decision Tree (DT). We selected MLP for its state-of-the-art performance, LR for its simplicity, and DT for their non-differentiability. Finally, we perform the black-box attacks on both attacker knowledge of the training data scenarios while increasing the computational faults rate, i.e., decreasing the supply voltage.

Figure~\ref{fig:reverse} shows the black-box attack (reverse-engineering) effectiveness of the two attack scenarios (attacker have and does not have access to victim's training data) while increasing the VOS-induced computational faults rate. The results show that using a \shmd with a $0.1$ fault rate makes the black-box attack substantially more difficult; the reverse-engineering effectiveness using MLP drops from $99.1\%$ to $75.5\%$ (around $24\%$ drop) when using the \textit{attacker training} set and from $99.2\%$ to $86.0\%$ (around $13.3\%$ drop) when using the \textit{victim training} set. Furthermore, the results show that the \shmds resilience to black-box attacks increases by increasing the computational faults rate, irrespective of the machine learning algorithm used to perform the attack. As seen from the results, reverse-engineering attacks become harder with VOS.

\begin{figure}[]
\includegraphics[width=\linewidth]{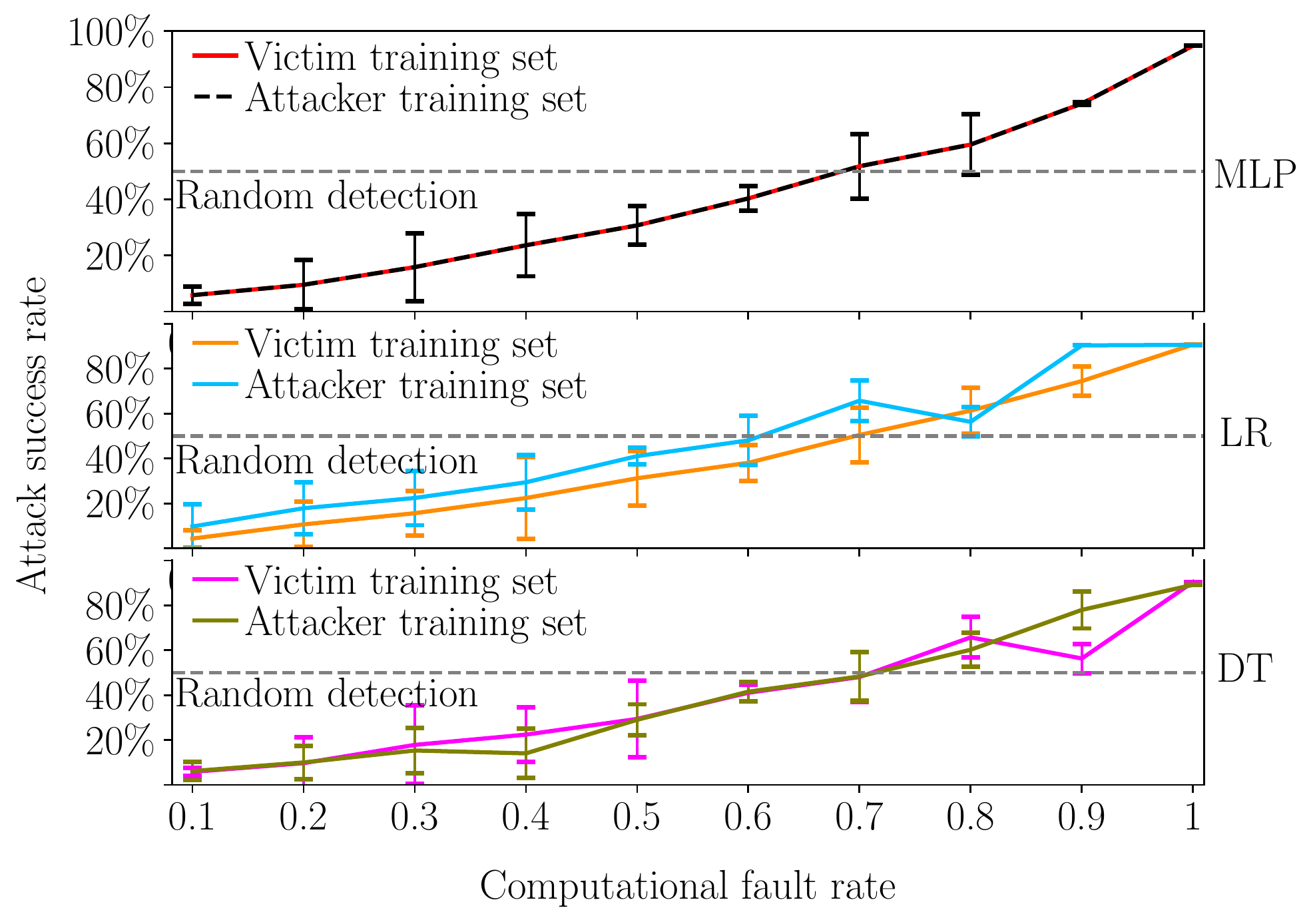} 
\centering
\caption{\shmds robustness against transferability of evasion attacks }
\label{fig:transferability}
\end{figure}

\noindent
\textbf{Transferability of attacks:} having a reversed-engineered model of the victim HMD, transferability is defined by the percentage of evasive malware designed to evade the reversed-engineered model that can also evade the victim HMD's detection, i.e., transfer to the defender model. Therefore, in this experiment, we generated evasive malware samples based on each of the reverse-engineered models of the \shmd (shown in Figure~\ref{fig:reverse}), and tested their success rate in evading the \shmd. Figure \ref{fig:transferability} shows the transferability results when attacking \shmds. The results show that the transferability of the attacks increases, i.e., the evasions are more successful, while increasing the computational faults rate. This can be explained by the impact of computational faults on the decision boundary. In fact, moving the decision boundary while keeping the classification reliable is an efficient defense against adversarial attacks in general \cite{def_appx}. However, aggressively increasing the computational faults rate results in a major accuracy drop that impacts the baseline model reliability, including its capacity to recognize evasive malware.
In addition to detection accuracy and transferability robustness, a comprehensive trade-off investigation should also include the robustness against reverse-engineering, which we discuss in Section \ref{sec:Disc}.




\section{Detection Performance Evaluation}\label{sec:perf}
After showing \shmds resilience against adversarial attacks, in this section, we evaluate \shmds performance. Specifically, we study the impact of VOS-induced computational faults on detection accuracy and detection speed.

\subsection{Detection accuracy}\label{subsec:shmd_accuracy}

This study examines the effect of VOS-induced computational faults on HMDs detection accuracy. Specifically, we evaluated the detection performance of \shmds under no attacks. Therefore, we used the \textit{testing} set to evaluate the detection accuracy of \shmd.

\begin{figure}[htp!]
\includegraphics[width=\columnwidth]{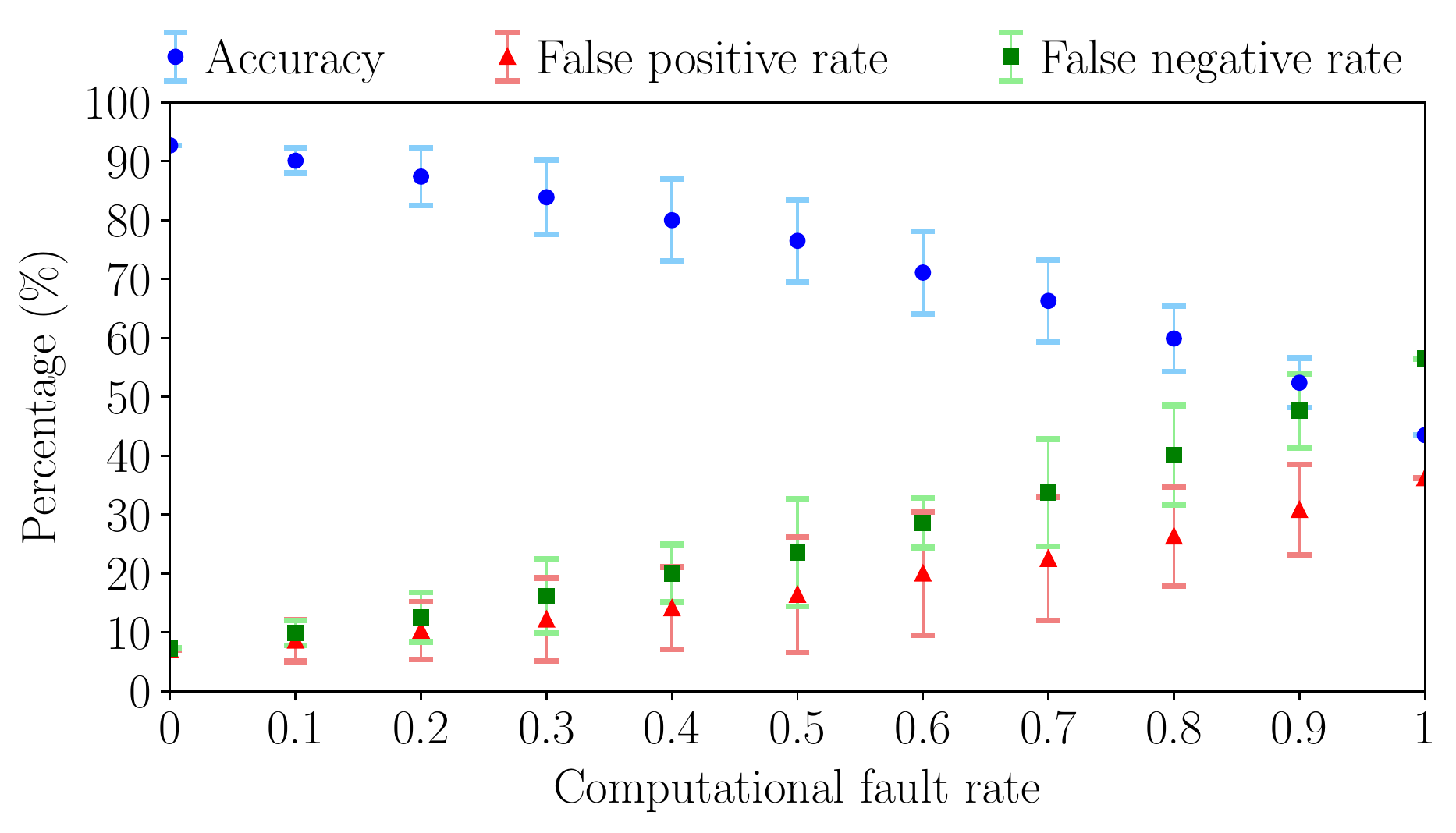} 
\centering
\caption{VOS effect on HMD's detection performance}
\label{fig:vs}
\end{figure}

Figure~\ref{fig:vs} shows the \shmd detection accuracy, false positive rate, \textcolor{black}{and false negative rate} while increasing the computational faults rate (scaling the voltage). \textcolor{black}{We performed 4-folds cross-validation and repeated each experiment 50 times to obtain representative results}. 
An interesting observation is that the standard deviation increases while increasing the VOS until a $0.5$ computational faults rate, and then it starts decreasing. Notice that the standard deviation represents the stochasticity that VOS adds to the output due to the non-deterministic decision boundaries. Figure~\ref{fig:vs} also shows that the accuracy degradation diverges logarithmically as the computational faults rate approaches 1; the relationship is not linear. The same observation also applies to the false positive rate and false negative rate (increases logarithmically as the computational faults rate approaches 1). This is a strong advantage from the defender perspective since adding more computational faults (specifically, until $0.5$ computational faults rate) would not significantly impact detection accuracy loss. For example, at $10\%$ computational faults rate, the detection accuracy of \shmds drops by around $2\%$ only. Furthermore, increasing the computational faults rate from $0.1$ to $0.4$ ($4\times$ increase in computational faults) would result in only $0.1\times$ detection accuracy loss.  

\subsection{Detection speed}\label{subsec:shmd_speed}
\begin{figure}[]
\includegraphics[width=\columnwidth]{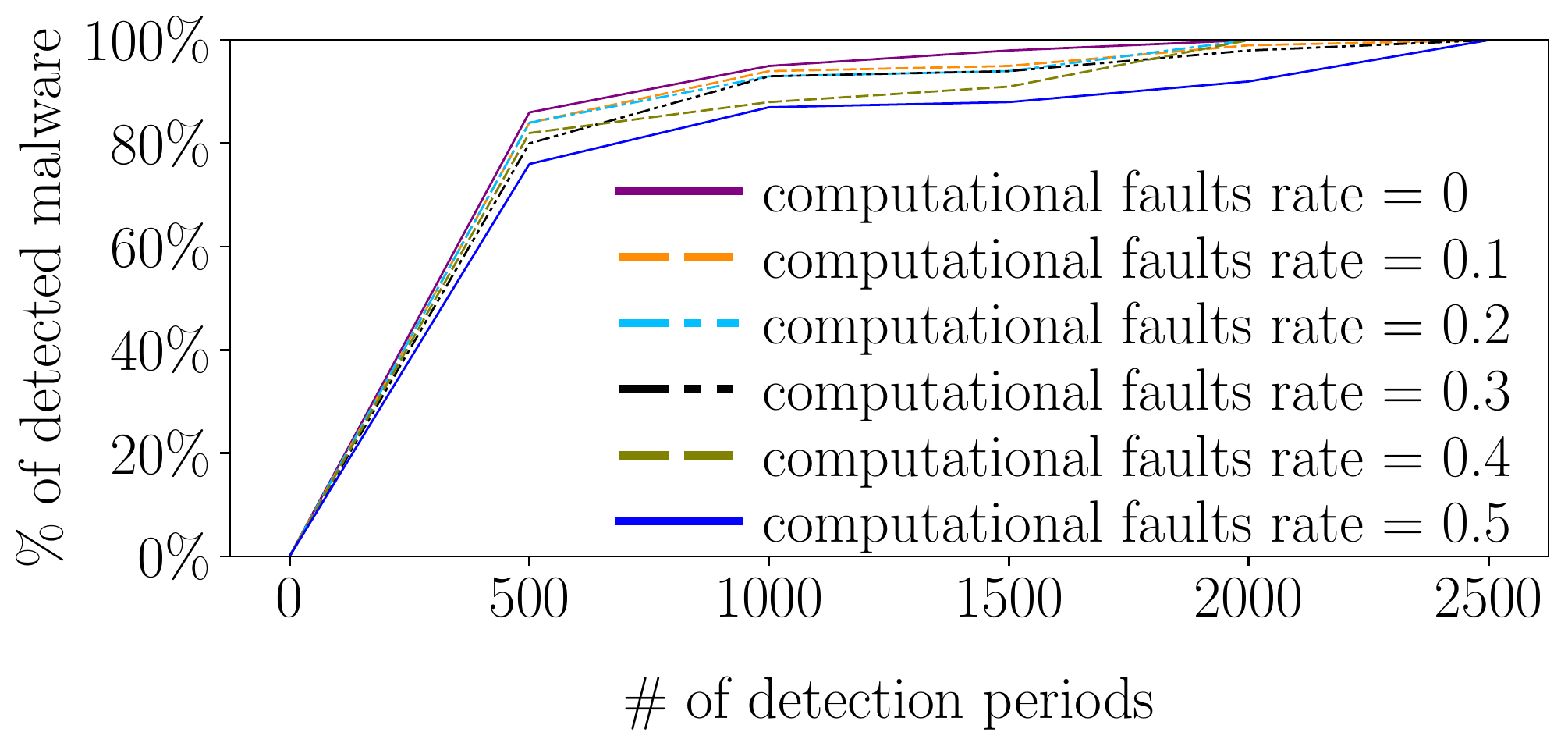} 
\centering
\caption{\shmd detection speed}
\label{fig:spdee}
\end{figure}

This study examines the effect of VOS-induced computational faults on HMDs detection speed. The detection speed is defined as the number of detection windows/periods it takes to detect a malware program since it started execution. 
Figure~\ref{fig:spdee} shows a cumulative probability distribution of the percentage of detected malware programs within a number of decision range for different computational faults rates. Although not inducing VOS computational faults (i.e., baseline HMD) outperforms any VOS-induced computational fault rate in this metric, the advantage is relatively small. For example, regular HMD detects $85.7\%$ of malware in $500$ periods or less, while at $0.1$ computational faults rate it can detect $84.4\%$  of malware and at $0.5$ computational faults rate it can detect $76.2\%$ of malware. On average, 
\shmd can detect malware within $243$, $240$, $236$,	and $234$ detection periods when the computational faults rate is $0$, $0.1$, $0.2$, and $0.3$ respectively.

\section{Implementation Overhead Evaluation}

\begin{figure}[h]
        \centering
         \subfloat[Hardware implementation]{
                \includegraphics[width=0.9\linewidth]{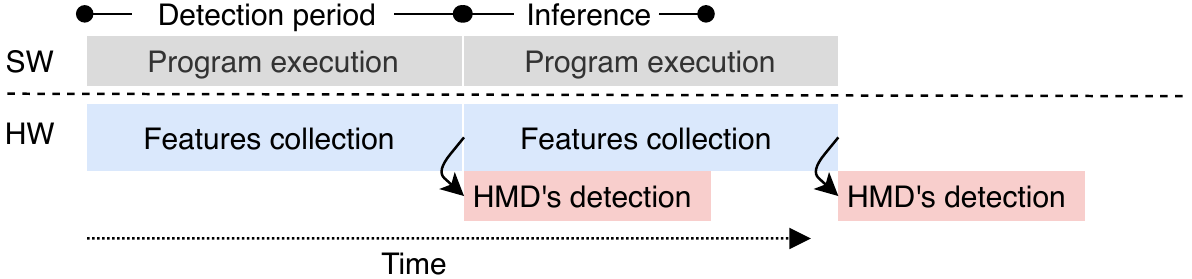}
                \label{subfig:hard}
        }
        \vspace{2mm}
        \subfloat[Software implementation]{
                \includegraphics[width=0.9\linewidth]{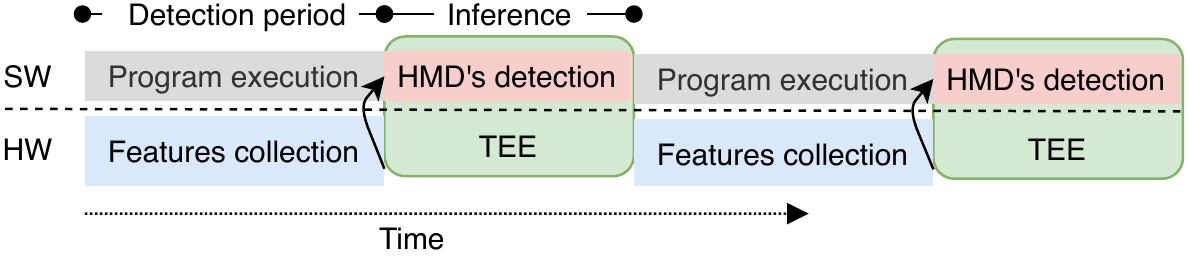}
                \label{subfig:soft}
        }
        \caption{HMDs implementations overview}
        \label{fig:HMDs_imp}
\end{figure}

HMDs are detectors that use low-level hardware features for detection. However, the inference part of the detection can be implemented in two main methods, namely hardware and software. In a hardware implementation, the HMDs' inference computation run on a co-processor implemented in the hardware and a monitoring unit to collect features is also implemented in the hardware. In contrast, for a software implementation, the HMDs collect features using either HPCs, if they are available, or a specially designed hardware unit to extract features, however, the HMDs run inference in a trusted execution environment (TEE), e.g., Intel SGX~\cite{costan2016intel}. An overview that shows where and when HMDs detection (inference) takes place during a program execution for both hardware and software implementations is shown in Figure~\ref{fig:HMDs_imp}. In this section, we show how \shmds can be implemented in both hardware and software. Furthermore, we compare \shmds implementations overhead to state-of-the-art randomization based solution for HMDs, RHMDs~\cite{khasawneh2017rhmd}. Basically, RHMDs is a randomization-based defense that switches between multiple models, i.e., decision boundaries, at run-time to perform malware detection. The RHMDs' resilience increases while increasing the number of models (base-detectors). Our results show that \shmds offer storage space, power, and hardware logic resource savings over RHMDs, and offer power savings over baseline HMD. Also, note that in all of our experiments, we compared \shmds to the most efficient RHMD structure, which uses only 2 base-detectors; this is not the most secure structure though, as the resilience of RHMDs increases while increasing the number of base-detectors\cite{khasawneh2017rhmd}. 

\subsection{Hardware-based \shmds}

\begin{figure}[]
\includegraphics[width=\columnwidth]{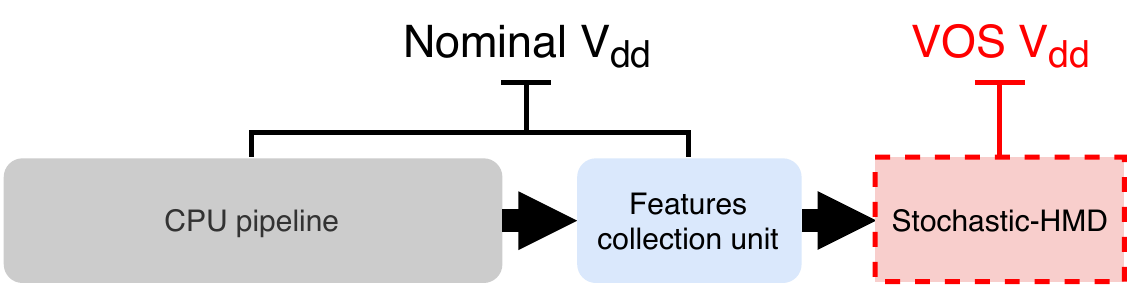} 
\centering
\caption{Architectural overview of hardware-based \shmds, where $Nominal\  V_{dd} >$ 
$VOS\ V_{dd}$}
\label{fig:imp_shmds}
\end{figure}

\noindent
\textbf{Implementation:} The general \shmds architecture view is depicted in Figure~\ref{fig:imp_shmds}. Compared to a baseline HMD, i.e., insecure HMD, the only difference is that the \shmd co-processor IP, i.e., that does the inference computations, is fed by a  supply voltage $VOS-V_{dd}$ that is lower than the $nominal-V_{dd}$ that feeds the rest of the system. Keeping the other components under nominal supply voltage is important to avoid unreliable overall behavior. \textcolor{black}{Furthermore, trusted control of VOS can be guaranteed by dedicating one of the several integrated voltage regulators (VRs) in the processor to the \shmd IP.  }





To assess resource utilization experiments, we implemented both \shmd and RHMD using Verilog. Specifically, we implemented the detection co-processor IP of both \shmd and RHMD. Figure~\ref{fig:comp_imp} shows the architectural view of our implementations. Notice that compared to an insecure HMD, the supply voltage of the co-processor IP of \shmd is scaled down, compared to $nominal-V_{dd}$,  without scaling the frequency (VOS). In contrast, RHMDs have to store multiple models and select one of them randomly at each detection.

\noindent
\textbf{Overhead:} We synthesized both \shmd and an RHMD (configured to have 2 and 4 base-detectors) on a Xilinx Zynq 7000 FPGA board. \textcolor{black}{The \shmd and RHMD base-detectors has the same MLP architecture described in Section~\ref{subsec:victimHMD}. The implementation consists of pipelined, fully unrolled processing units consisting each of a MAC element and a MUX-based Relu function.} We evaluate the implementation overhead using the following metrics:



\begin{figure}[h]
\includegraphics[width=\columnwidth]{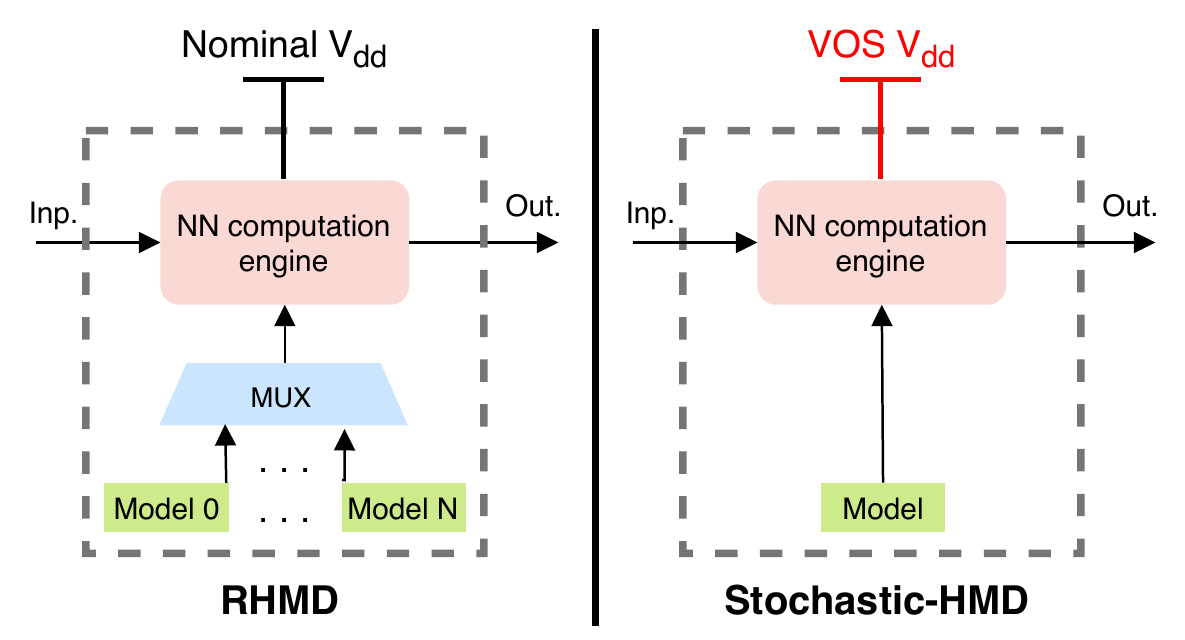} 
\centering
\caption{\shmds and RHMDs inference implementation}
\label{fig:comp_imp}
\end{figure}

\noindent
\textit{(1) Area:} 
 The normalised resource utilization overhead compared to an insecure HMD implementation is shown in Table~\ref{tbl:ru_overhead}. We observe that \shmds consumes around $3 \times$ less Look Up Tables (LUTs) than the baseline RHMDs, i.e., with only two base-detectors (models). In addition, we implemented an RHMD with 4 base-detectors and observed that this configuration consumes $4.87 \times$ and $3.67\times$ more LUTs and BRAMs, respectively. The number of DSP slices is comparable since RHMD can be implemented by reusing computation elements between different models.

\begin{table}[h!]
\centering
\small
\caption{Normalized resource utilization overhead of \shmds and N-RHMDs compared to insecure HMDs. N denotes the number of base-detectors (models) in the RHMD configuration}
\label{tbl:ru_overhead}
\begin{tabular}{lcc}
 \hline
                                        
             \begin{tabular}[c]{@{}c@{}}\textbf{}\end{tabular}  &
             \begin{tabular}[c]{@{}c@{}}\textbf{LUTs}\end{tabular}  &
             \begin{tabular}[c]{@{}c@{}}\textbf{BRAMs}\end{tabular}  \\ \hline
 
  2-RHMD &  $3.06\times$    & $1.89\times$     \\ \hline
 4-RHMD &  $4.87\times$    & $3.67\times$\\ \hline
  \shmd &  $1\times$    & $1\times$     \\ \hline
\end{tabular}
\smallskip
\end{table}

\noindent

\begin{figure}[htp!]
\includegraphics[width=\columnwidth]{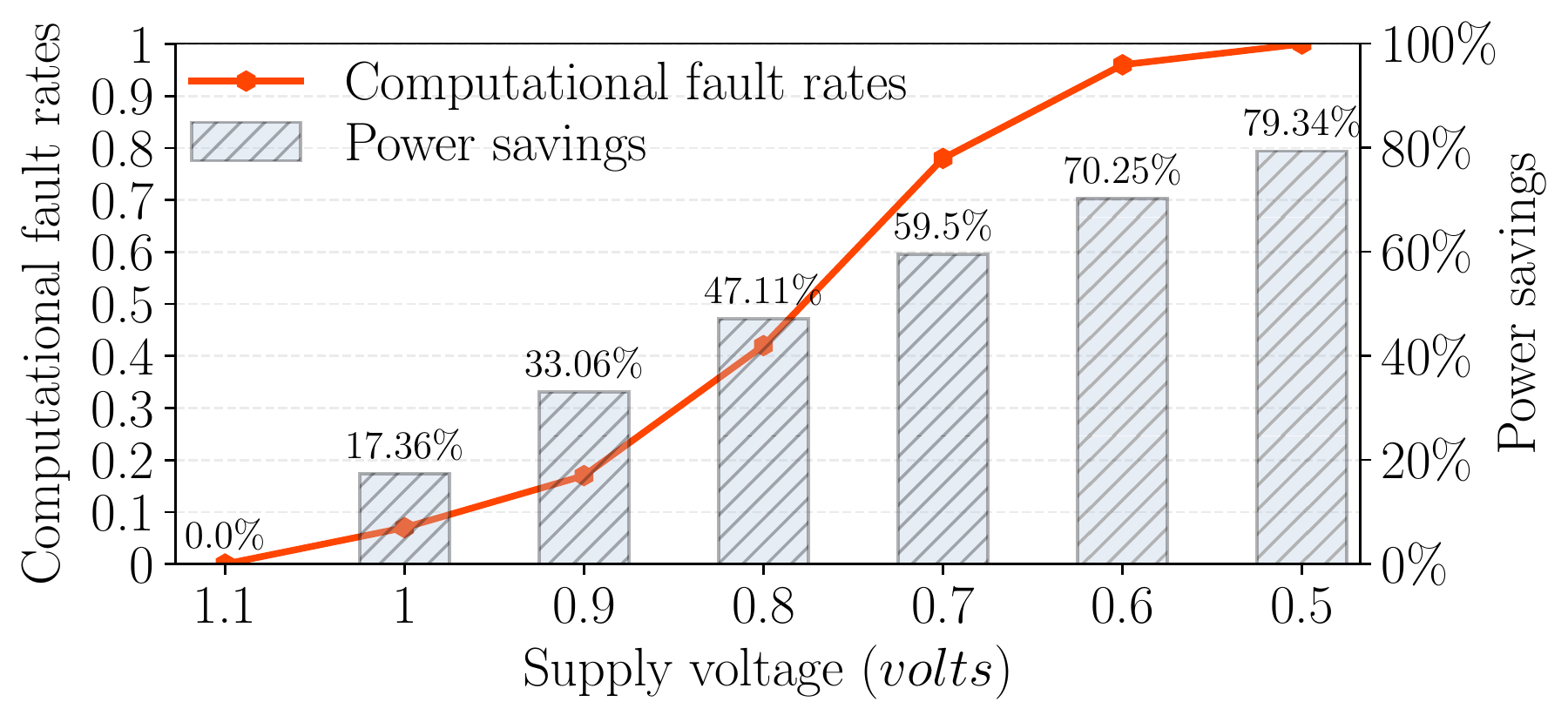} 
\centering
\caption{Multiplier power saving under VOS}
\label{fig:pow}
\end{figure}

\label{subsec:pow_sav}
\noindent
\textit{(2) Power consumption:}
There are no interfaces or configuration parameters to scale the voltage in FPGAs boards. 
Therefore, we implement a multiplier at $90~ nm$ technology with PTM \cite{PTM} using Keysight Advanced Design System (ADS) simulation platform and run Monte Carlo simulations to evaluate the circuit computational faults rate under VOS with thermal and process variation. This choice is due to the high impact of multipliers in terms of resource utilization compared to adders in an artificial neural network. Figure~\ref{fig:pow} shows the dynamic power saving corresponding to the multiplier accuracy loss. 
While these are expected results, they emphasize interesting trade-offs that could be obtained from \shmds. For instance, under a computational faults rate of $0.4$, the accuracy drops by around $10\%$, and we save $47\%$ of power consumption.  


\subsection{Software-based \shmds}

\noindent
\textbf{Implementation:} In a software-based HMD implementation, regardless of the feature collection, the HMDs detection, i.e., inference, will be executed in a TEE to prevent a compromised operating system from hijacking/corrupting the detection, as shown in Figure~\ref{subfig:soft}. 

To implement a software-based \shmd, we have to find a way to scale the voltage. In particular, the voltage needs to be over-scaled directly after entering the TEE and scaled back to the nominal voltage just before exiting the TEE. Luckily, the reverse-engineering efforts on modern CPUs revealed that chip-level dynamic voltage scaling can be controlled from software through the undocumented Model-Specific Registers (MSRs)~\cite{Miha_Guide,RightMark_RMClock}. It has been shown that VOS can be applied using MSRs to induce statistical computational faults in modern CPUs and undermine the system's security~\cite{Murdock2019plundervolt}. In contrast, our work utilizes such stochastic computations to improve the security of HMDs. \textcolor{black}{Furthermore,
the trust control of VOS can be supported using two mechanisms; (1) tampering detection: by simply checking the voltage value during detection, and (2) control: current processors have several integrated VRs, thus, the OS can grant exclusive control of the StochasticHMD’s VR to the HMD’s enclave during detection. }


Therefore, we used an Intel Broadwell processor (model number i7-5557U) running an Ubuntu 16.04 with stock Linux v4.15. We scale the voltage using the MSR register $\mathtt{0x150}$. Specifically, we set the $\mathtt{plane\ idx}$ bits to $\mathtt{0}$ to scale the core's voltage only, and used the $\mathtt{offset}$ bits to scale the voltage, which are used to scale the voltage by the specified offset in $mV$~\cite{Murdock2019plundervolt}. Moreover, the core frequency was kept at 3.1 GHz at the time of the experiments.


\begin{table}[h!]
\centering
\small
\caption{Number of multiplications required to observe a single computational fault vs. necessary voltage offset from $nominal- V_{dd}$ on i7-5557U at 3.1 GHz}
\label{tbl:dvfs}
\begin{tabular}{rc}
 \hline
             \begin{tabular}[r]{@{}r@{}}\textbf{Iterations}\end{tabular}  &
             \begin{tabular}[c]{@{}c@{}}\textbf{Offset}\end{tabular}  \\ \hline
               $1,000,000,000$ &  $-107mV$    \\ \hline
                 $100,000,000$ &  $-108mV$    \\ \hline
                  $10,000,000$ &  $-109mV$    \\ \hline
                   $1,000,000$ &  $-112mV$    \\ \hline
                     $500,000$ &  $-114mV$    \\ \hline
                     $100,000$ &  $-115mV$    \\ \hline
               
\end{tabular}
\smallskip
\end{table}

Table~\ref{tbl:dvfs} shows the number of multiplications required to observe a single computational fault for a specific voltage in our experimental setup. The results show that the probability of computational faults occurrence increases while scaling the voltage. Two challenges are important to mention here: (i) high VOS results in crushing the whole system, and (ii) even with low VOS and low error rates, we were not able to identify faults locations, i.e, memory or computations. For these reasons, we evaluated the software implementation in terms of memory space, execution time, and power.


\noindent
\textbf{Overhead:} We evaluate \shmds software-based implementation overhead against RHMDs. Moreover, we used the same features described in Section~\ref{sec:data} in our experiments since we are evaluating the HMD's inference overhead regardless of the feature collection approach. 

\noindent
\textit{(1) Storage space:} 
 RHMDs store multiple HMDs models, i.e., base-detectors, to be able to switch between them. In contrast, in a \shmd only one model needs to be stored. Therefore, storage savings of \shmds over RHMDs can be measured using the following equation: 
 \vspace{-7mm}

\begin{equation} 
\textstyle
Storage\ savings = \frac{\#\ of\ base\ detectors\ in\ RHMD - 1}{\#\ of\ base\ detectors\ in\ RHMD}
\end{equation}

\noindent
For example, \shmd storage saving over an RHMD, with only 2 base-detector, is 50\%. This saving not only reduces the storage space in memory, but also the pressure on the CPU resources, such as the caches and the bus; thus, reducing the overhead on the system overall performance. For example, in our experiments, every HMD takes 71 KB of memory, while the L1 cache size in Intel's latest CPU generation, i.e., Tiger Lake, is 32 KB.

\noindent
\textit{(2) Inference time:}
 We ran 1000 detection, i.e., inferences, for both \shmd and RHMD (with 2 base-detectors).
 We noticed an average of $8\%$ overhead of RHMD over \shmd.  
 The inference time of RHMDs is higher owing to its additional task of randomly selecting a model from its set of base models; such random model selection also has an impact on L1 cache eviction. We noticed that scaling the voltage has no effect on the inference time. This is because VOS has no effect on the cycle time since we are only scaling the voltage but not the frequency. 

\begin{figure}[]
\includegraphics[width=\columnwidth]{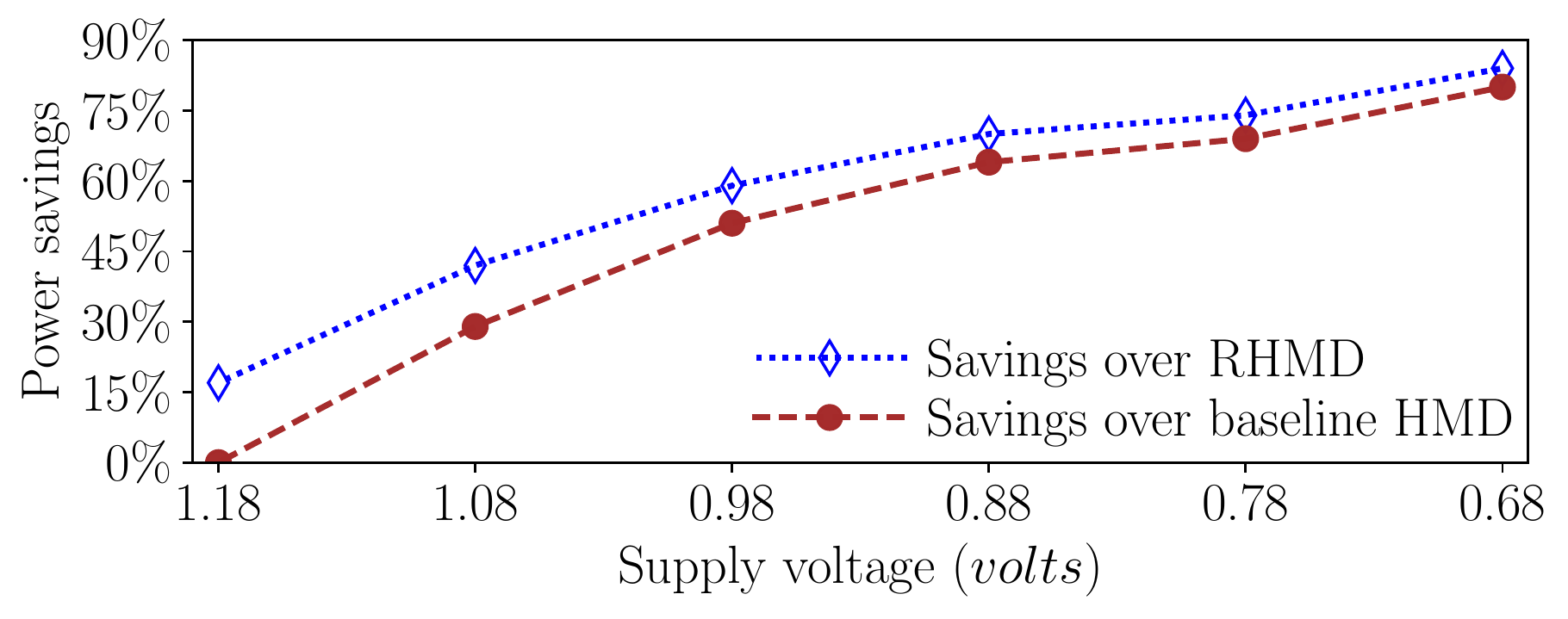} 
\centering
\caption{Power savings of \shmd over baseline HMD and RHMD. }
\label{fig:model_power}
\end{figure}

\noindent
\textit{(3) Power consumption:}  We ran 1000 detection, i.e., inferences, for both \shmd and RHMD (with 2 base-detectors). We use Intel's Power Gadget~\cite{kim2014intel} to measure the power consumption and report the average consumed power per inference.
Figure~\ref{fig:model_power} shows the power savings for voltages ranging from $1.18~V$ (the nominal voltage to $0.68~V$ with a voltage step size of $0.1 V$. Results show over $75\%$ in power saving compared to RHMD achieved by \shmd under $40\%$ voltage scaling.


\section{Theoretical Basis}
We provide a theoretical basis for the resilience of \shmds. Mainly, we show that \shmds is inherently more difficult to be reverse-engineered than a conventional HMD.
We base our analysis on the probably approximately correct (PAC) learnability theory \cite{pac}.
\subsection{Deterministic HMD Learnability}

Consider a learning system that corresponds to the defender implementing a conventional HMD. Consider also an attacker attempting to design a reverse-engineered learning system. This latter proceeds to querying the defender classifier to learn its behavior and transfer evasion attacks to it. 
Let $H$ be the class of possible classifiers a learning system considers and $P$ the probability distribution over data instances $(x,y)$, where $x$ represents an input feature vector, and $y$ a label in $\{0,1\}$.
Let be $y= \overline{h}(x)$ the ground truth, i.e., a deterministic function that gives the correct label of an input $x$. For any $h \in H$, let $e(h)=\operatorname{Pr}_{x \in P}[h(x) \neq \bar{h}(x)]$ be the error of $h$ w.r.t. $P$ and $e_{H}=\inf _{h \in H} e(h)$ the smallest achievable error by any function $h$. Let $D= \{ (x_1,y_1), ..., (x_m,y_m)\}$ be a training data set of size $m$ that follows $P$ and $\mathscr{D}$ be the set of all possible data sets $D$. A learning algorithm is a function $L: \mathscr{D} \rightarrow H$ which produces a classifier $\hat{h} \in H$ given D.

\begin{definition}
A hypothesis class $H$ is learnable if $\exists L$, a learning algorithm for $H$ such that $ \forall \varepsilon, \delta \in [0,\frac{1}{2}]$ and distribution $P$, it exists a training sample size $m_0 \left(\varepsilon, \delta \right)$, s.t. $\forall {m \geq m_0\left(\varepsilon, \delta \right)}, \operatorname{Pr}_{D \in \mathscr{D}}\left[e(L(D)) \leq e_{H}+\varepsilon\right] \geq 1-\delta$, i.e., $L$ will with at least probability $(1-\delta)$ output a hypothesis $\hat{h} \in H$, whose error in $P$ is almost $(e_H + \varepsilon)$. $H$ is \emph{efficiently learning} if $m_0(\varepsilon, \delta)$ is polynomial in $\frac{1}{\varepsilon}$ and $\frac{1}{\delta}$, and $L$ runs in time polynomial in $m,\frac{1}{\varepsilon} and \frac{1}{\delta} $.
\end{definition}
This definition implies that a hypothesis class $H$ is efficiently learnable if one can get with high probability an approximately optimal hypothesis from $H$ given a polynomial number of samples. The error bound $e_H + \varepsilon$ for an approximately correct classifier $\hat{h} \in H$ consists of two components: \textbf{(i)} $\varepsilon$, becomes arbitrarily small and hence $e(\hat{h})$ approaches $e_H$ when the number of training samples increases polynomially w.r.t. $\frac{1}{\varepsilon}$; and \textbf{(ii)} $e_H$ which depends on the learning bias about $H$ \cite{pac}. We observed the implications of this result in Table \ref{tbl:vhmd_per}.

This learnability definition applies to the defender and the attacker, as soon as the attacker has access to enough data to train his proxy model as observed in Section \ref{sec:BBA}. In the next subsection, we study the learnability of the attacker model in the case of a \shmd defender.

\subsection{\shmd Learnability}

In this section we consider a defender that uses a \shmd whith the possibility to tune the VOS. Let be $y= \overline{h}(x)$ the ground truth in a class of possible classifiers $H$ and a data distribution $P$ over $(x,y)$ corresponding to input features and labels respectively. Let $e(h)=\operatorname{Pr}_{x \in P}[h(x) \neq \bar{h}(x)]$ be the expected error of $h \in H$ w.r.t. $P$ and $e_{H}=\inf _{h \in H} e(h)$ is the smallest error achievable by any function $h \in H$.

Let $H_i , i\in [0,n]$ be a hypothesis class where $i$ represents the VOS index and $\hat{h}_i \in H_i $ a \shmd classifier that corresponds to the $i^{th}$ voltage scaling level such that: \textbf{(i)} $\hat{h}_0 = h$ ($i.e.$ nominal supply voltage), and \textbf{(ii)} the higher the index $i$, the more aggressive the VOS. 

Because of the stochastic aspect of $\hat{h}_i$, the classifier $\hat{h}_i$ has different instances; an instance $k$ of $\hat{h}_i$ is denoted $\hat{h}^{k}_{i}$. We denote $H^k_i$ the class of possible classifiers the \shmd considers under the $i^{th}$ VOS level.
We define $\triangle_{k, r}=\operatorname{Pr}_{x \in P}\left[\hat{h}^{k}_{i}(x) \neq  \hat{h}^{r}_{i}(x)\right]$ that measures the difference between two instances $k$ and $r$ of the same \shmd at a given VOS level $i$. Notice that $\triangle_{k, r}$ is a function of $i$; the more aggressive the VOS, the higher $\triangle_{k, r}$ is likely to be.  Consider a space of policies defined by the VOS index $i$ and parametrized by $p^k_{i} \in [0,1]$ with $\Sigma_{k} p^k_{i}=1$ where $\hat{h}^{k}_i \in H^k_i $ occurs with a probability  $p^k_{i}$.

Let $\vec{p}$ denote the corresponding probability vector. Then, a policy $\vec{p}$ induces a distribution $Q_{\vec{p}}$ over $(x, z), \text { where } z=\hat{h}^k_{i}(x)$ with probability $p^k_{i}$. For a given VOS level, the defender will incur a baseline error rate of $e_{\vec{p}}(\hat{h})=\operatorname{Pr}_{x \in Q_{\vec{p}}}\left[\hat{h}^{k}_{i}(x) \neq \bar{h}(x)\right]=\sum_{i} p^k_{i} e\left(\hat{h}^{k}_{i}\right)$. 

Now suppose the attacker observes a sequence of queries from $Q_{\vec{p}}$ for a given voltage $i$, and tries to efficiently learn the hypothesis class $H=\cup_{k} H^{k}_{i}$ to reverse engineer the model. 

For any $h \in H$, let $e_{\vec{p}}(h)=\operatorname{Pr}_{x \in Q_{\vec{p}}}\left[h(x) \neq \hat{h}^k_{i}(x)\right]=\sum_{i} p^k_{i} e(h)$, the expected error of $h$ w.r.t. $Q_{\vec{p}}$, and we define $e_{\vec{p}, H}=\inf _{h \in H} e_{\vec{p}}(h)$ as the smallest error achievable by any function $h \in H$ under a policy $\vec{p}$. Notice that Definition 1 naturally extends to the stochastic setting: in particular, the distribution $P$ becomes $Q_{\vec{p}}$ and the error bound $\left(e_{H}+\varepsilon\right)$ becomes $\left(e_{\vec{p}, H}+\varepsilon\right)$.

\begin{theorem}
\label{theo:1}
Suppose that each $H^k_i$ is efficiently learnable, and $\hat{h}^k_i \in H^k_i$ be the classifier learned from these classes by a defender, and implemented using \shmd inference. Let $e(\hat{h}^k_i)$ the corresponding error rate of the hypothesis $\hat{h}^k_i$. Then any distribution $Q_{\vec{p}}$ over (x,z) can be reverse engineered, with $e_{\vec{p}, H}$ bounded by $\min _{k} \sum_{p \neq r} p
^k_{i} \triangle_{k,r} \leq e_{\vec{p}, H} \leq 2\left(\max _{k} e\left(\hat{h}^k_{i}\right)\right)$. \footnote{This theorem is formed by combining Theorem 2.2 and Corollary 2.3 (with detailed
proofs) from \cite{theorem}.}
\end{theorem}

This theorem reveals two observations:
\textbf{(i)} even with stochastic inference, reverse-engineering is still possible as long as the VOS doesn't affect the classification accuracy  - that means that $\max _{k} e\left(\hat{h}^k_{i}\right)$, the maximal error at a given VOS index $i$, is arbitrarily small. This is the case of a low magnitude of $\triangle_{k,r}$ and a non-aggressive voltage scaling.\\ 
\textbf{(ii)} The attacker’s error depends directly on
the stochasticity of the classifiers, which can be significant if at
least some of the runs are often not very accurate. This phenomenon can be observed with the aggressive VOS.
According to the error bound $\left(e_{\vec{p}, H}+\varepsilon\right)$, even
though $\varepsilon$ becomes arbitrarily small as the number of queried samples increases, the defender will inevitably suffer from an error caused by $e_{\vec{p}, H}$. 
This error can be high; for example, when scaling the voltage to $0.7 V$, the error $e_{\vec{p}, H}$ is in $[0.35, 0.45]$. In contrast, graceful VOS, such as a supply voltage of $1 V$, the error can be negligible. 

Theorem \ref{theo:1} also suggests a trade-off between the accuracy
of the defender under no reverse-engineering vs. the susceptibility
to being reverse-engineered: using low-accuracy but high-stochasticity classifiers allow the defender to enhance the robustness of the classifier, but also degrade its performance.
\section{Discussion}
\label{sec:Disc}

\begin{figure}[!ht]
\includegraphics[width=\columnwidth]{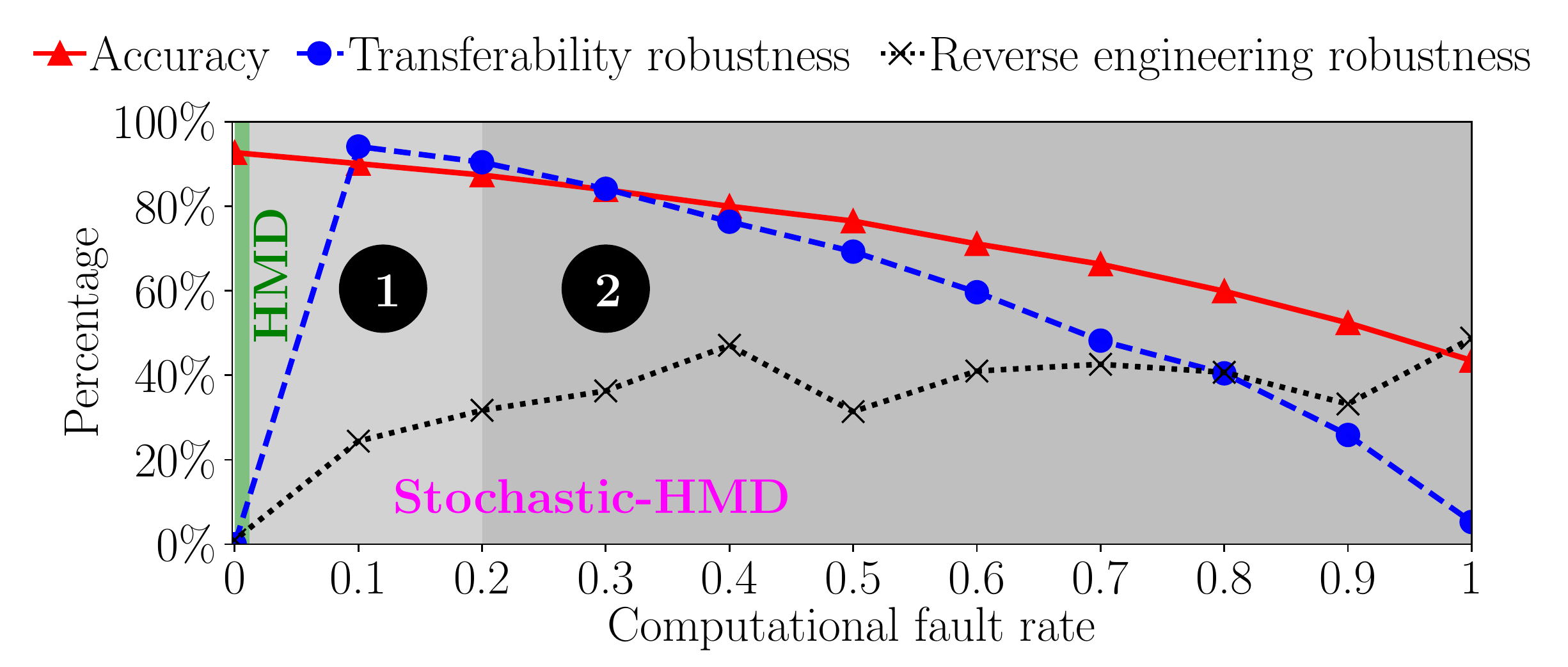} 
\centering
\caption{\shmd trade-off}
\label{fig:tradeoff}
\end{figure}

This work has unprecedentedly shown that a potential robustness enhancement technique of HMDs can be implemented in a practical and efficient manner using VOS. The results show that \shmds offer controllable randomization that hardens HMDs against black-box adversarial attacks. In particular, VOS-induced stochastic noise makes the model reverse-engineering harder. This property goes crescendo with increasing error rates. On the other hand, the robustness to evasive malware transferability is higher under "soft" VOS; increasing error rates reduces the defender's robustness to this attack setting. For an overall efficient HMD, these security aspects need to be considered under the baseline accuracy constraint, i.e., the accuracy without evasive malware. An overview of these different trends and the possible tradeoff is shown in Figure \ref{fig:tradeoff}. Specifically, Figure \ref{fig:tradeoff} shows the detection accuracy, transferability robustness, i.e., percentage of evasive malware created using the proxy model that fails to evade victim HMD's detection, and reverse-engineering robustness, i.e., loss in reverse-engineering effectiveness ($100 -$ reverse-engineering effectiveness). From the figure, it is clear that a fault rate that exceeds $0.2$ (area $\numcircledtikz{2}$) would no longer be practical. However, as shown in area $\numcircledtikz{1}$, trade-offs between security and performance could be reached with low error rates, which can considerably increase the defender HMD reverse-engineering robustness, transferability robustness (achieve more than $94\%$, even with effectively reverse-engineered victim HMD), with negligible accuracy loss, along with around $15\%$ power saving.

While this work focuses on HMDs security, we believe that this finding could be efficient beyond this use case and apply to generic machine learning applications, which we will investigate in our future work. 

\section{Related Work}\label{sec:related}
In this section, we discuss defense approaches against adversarial attacks and some important recent results in this domain. Various approaches have been proposed to help defend against adversarial attacks. Therefore, we organize this section based on the general groups of defenses.

\noindent
\textbf{Adversarial training:} 
Adversarial training is one of the most explored defenses against adversarial attacks~\cite{samangouei2018defense, tramer2017ensemble, khasawneh2017rhmd}. The main idea is to train the model using both the training data set and evasive samples. As a result, the trained model will classify evasive samples with higher accuracy. Nonetheless, adversarial training is much more computationally intensive than training a model on the training data set only, because generating evasive samples needs more computation and model fitting while including evasive samples is more challenging (takes more epochs)~\cite{tramer2017ensemble}. Moreover, adversarial training is not effective when the attacker uses a different attack strategy than the one used to train the model~\cite{samangouei2018defense}.
In addition, it was shown that it does not defend against black-box attacks~\cite{khasawneh2017rhmd}.

\noindent
\textbf{Input Preprocessing:}
Input preprocessing depends on applying transformations to the input to remove the adversarial perturbations~\cite{ das2017keeping, osadchy2017no}. Examples of transformation are denoising auto-encoders~\cite{gu2014towards}, the median, averaging, and Gaussian low-pass filters~\cite{osadchy2017no}, and JPEG compression~\cite{das2017keeping}. However, these are only applicable for image classification but not for malware detection. In addition, preprocessing every input requires additional computation. Furthermore, it was shown that this group of defenses is insecure~\cite{chen2019towards}; if the attacker knows the specific used transformation, he can take this into account when creating the evasive sample.

\noindent
\textbf{Gradient masking:}
Gradient masking relies on applying regularization to the model to make its output less sensitive to input perturbations. Papernot et al. proposed defensive distillation~\cite{papernot2016distillation}, by distilling knowledge out of a large model to train a compact model that is more robust to adversarial perturbations. Nayebi and Surya~\cite{nayebi2017biologically} used saturating networks, which show more robustness to adversarial noise. Nonetheless, these solutions fail to solve the problem of evasion since they just make the construction of white-box evasive samples more difficult. In addition, they suffer from black-box attacks~\cite{papernot2017practical}. Wang et. al. proposed defensive dropout~\cite{wang2018defensive} by dropping some of the units (e.g., neurons) at training and inference time. However, it requires re-training,
which is already prohibitive in many cases.

Most related to our work is RHMD~\cite{khasawneh2017rhmd}. RHMDs is an HMD organization that uses multiple diverse detectors and switches between them unpredictably. Detectors built in this fashion are resilient to reverse-engineering (black-box attacks), and thus, creating evasive malware becomes harder since the attacker does not have access to the victim's model. However, RHMDs have multiple limitations: (1) they require massive data sets with diverse features to be able to build multiple diverse detectors, and (2) they have a high hardware implementation complexity and substantial software implementation overhead compared to \shmds since multiple detectors need to be implemented.



\section{Conclusion}
We propose \shmds that unprecedentedly leverage VOS to defend against adversarial attacks. We show that such detectors can prevent the adversary from having reliable access to their detection model (i.e., resist reverse-engineering) and reduces the success rate of transferability attacks. We show empirically and theoretically that their resilience to the attacks can be achieved with low accuracy cost. Moreover, our proposed approach offers additional benefits compared to other adversarial defenses approaches, such as energy savings and easier deployment.

\bibliographystyle{IEEEtranS}
\bibliography{acm_ref}

\end{document}